\documentclass[prl,aps,showpacs,epsf,superscriptaddress,twocolumn]{revtex4}

\def\be{\begin{equation}}
\def\ee{\end{equation}}

\usepackage{bm}
\usepackage{amsmath,epsfig}
\usepackage{mathtools}
\usepackage{amsfonts}
\usepackage{color}
\usepackage[usenames,dvipsnames]{xcolor}
\usepackage{mathrsfs}

\newcommand{\bea}{\begin{eqnarray}}
\newcommand{\eea}{\end{eqnarray}}
\newcommand{\bi}{\begin{itemize}}
\newcommand{\ei}{\end{itemize}}

\newcommand{\rr}{{\bf r}}

\newcommand{\up}{\uparrow}
\newcommand{\down}{\downarrow}

\newcommand{\Br}{\mathcal{B}}
\newcommand{\Vr}{\mathcal{V}}

\newcommand{\Dr}{\mathcal{D}}

\newcommand{\Gz}{G_0} 
\newcommand{\Gaz}{\Gamma_0} 
\newcommand{\bl}{}

\newcommand{\beginsupplement}{%
        \setcounter{table}{0}
        \renewcommand{\thetable}{S\arabic{table}}%
        \setcounter{figure}{0}
        \renewcommand{\thefigure}{S\arabic{figure}}%
     }

\begin{document}

\title{
Resummation of diagrammatic series 
with zero convergence radius
\\for strongly correlated fermions 
}

\author{R. Rossi}
\altaffiliation{Present address: 
Center for Computational Quantum Physics,
The Flatiron Institute, New York, USA. 
}
\affiliation{Laboratoire de Physique Statistique, Ecole Normale Sup\'erieure - Universit\'e PSL, CNRS, Sorbonne Universit\'e, Universit\'e Paris Diderot, Paris, France}

\author{T. Ohgoe}
\affiliation{Department of Applied Physics, University of Tokyo,
7-3-1 Hongo, Bunkyo-ku, Tokyo 113-8656, Japan}

\author{K. Van Houcke}
\affiliation{Laboratoire de Physique Statistique, Ecole Normale Sup\'erieure - Universit\'e PSL, CNRS, Sorbonne Universit\'e, Universit\'e Paris Diderot, Paris, France}

\author{F. Werner}
\affiliation{Laboratoire Kastler Brossel, Ecole Normale Sup\'erieure - Universit\'e PSL, CNRS, Sorbonne Universit\'e, Coll\`ege de France,  Paris, France}

\date{\today}

\begin{abstract}
We demonstrate that summing up series of Feynman diagrams can yield unbiased accurate results for strongly-correlated fermions even when the 
convergence radius vanishes.
We consider the unitary Fermi gas,
a model of non-relativistic fermions in three-dimensional continuous space.
Diagrams are built from partially-dressed or fully-dressed propagators of single particles and pairs.
The series is resummed by a conformal-Borel transformation that
incorporates
the large-order behavior and the analytic structure in the Borel plane, which are found by the instanton approach.
We report
highly accurate numerical results for the equation of state in the normal unpolarized regime, and 
reconcile experimental data with the theoretically conjectured 
fourth virial coefficient. 
\end{abstract}
\maketitle

Feynman diagrams are a powerful computational tool and have led to an impressive list of important {\it approximate} results in various branches of physics.
But is it possible to make {\it accurate} predictions by summing up Feynman diagrams?

The answer is certainly yes if the coupling constant is small.
The most famous example is
quantum electrodynamics. 
Dyson argued that
the vacuum becomes unstable at negative fine-structure constant
and hence
the
 convergence radius should be zero~\cite{DysonCollapse}.
Nonetheless, thanks to the smallness of the coupling constant, 
the diagrammatic series behaves  as a convergent series for all practical purposes, 
leading to the most stringently tested physical theory~\cite{Kinoshita,CladeAlpha,GabrielseGmoins2}.

But what about strongly correlated theories?
In the pioneering work~\cite{LeGuillouZinnExposantsPRL,LeGuillouZinnExposantsPRB,GuidaZinn}
critical exponents were accurately computed
by summing up Feynman diagrams
in the strongly correlated regime of
$\phi^4$ theory. 
The problem of zero convergence radius was overcome by computing the large-order asymptotic behavior~\cite{Lipatov,BrezinLipatov} and using it to build an appropriate resummation technique based on a conformal-Borel transformation. 

For fermions
on a lattice, it is known in the mathematical physics literature that the convergence radius of diagrammatic series is non-zero in some part of the phase diagram~\cite{Mastropietro,MastropietroGraphene,MastropietroWeyl}.
In recent years, this allowed to obtain 
controlled thermodynamic-limit results in 
correlated regimes by summing up convergent diagrammatic series to high enough order using diagrammatic Monte Carlo~\cite{VanHoucke1,VanHouckeEPL,KulaginPRL,DengEmergentBCS,MishchenkoProkofevPRL2014,GukelbergerPwave,HuangPyro,IgorDirac,WuPseudogap},
where  the fermionic sign plays a
very different role 
than in conventional Quantum Monte Carlo~\cite{RossiComplexity}.

In this Letter, we report 
high-precision
results obtained by summing up Feynman-diagram series
for a strongly-correlated continuous-space fermionic theory
with zero convergence radius.
Specifically, we consider non-relativistic spin-1/2 fermions in three space
dimensions with contact interaction
-- a model 
which accurately describes 
ongoing ultracold atom experiments
and is also relevant to neutron
matter~\cite{RevueTrentoFermions,ZwergerBook,SylEOS,KuEOS,SalomonSFMixture,EsslingerQPC,RoatiJosephson,ZwierleinSolitonCascade,GrimmUltrafast,
GezerlisNeutronsReview,StrinatiUrbanReview}.
We derive the large-order asymptotic behavior of the diagrammatic series,
and we give mathematical arguments and numerical evidence for the resummability of
the series by a specifically designed conformal-Borel transformation that
incorporates the large-order behavior and the knowledge of the analytical structure
standing behind the series.
Combining this new resummation method with 
diagrammatic Monte Carlo evaluation up to order $9$,
we obtain
new results for the equation of state {\bl in the normal phase}, which agree with the ultracold-atom
experimental data from \cite{SylEOS,KuEOS}, except for the 4-th virial
coefficient for which our data  point to the theoretically conjectured value
of~\cite{EndoCastinConjecture}.

\begin{figure}
\includegraphics[width=1.0\columnwidth]{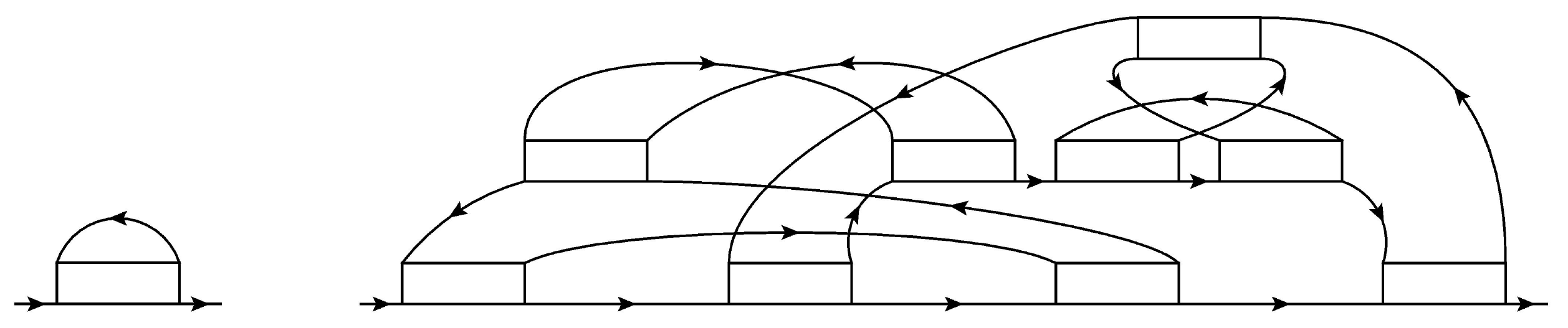}
\caption{First-order diagram (left) and an example of 9th-order diagram (right) for the single-particle self-energy. 
Each line represents a single-particle propagator while each box is a bosonic pair propagator. 
\label{fig:diag}}
\end{figure}

In order to have a well-defined diagrammatic framework for the contact interaction in continuous space,
it is necessary to incorporate exactly the two-particle scattering problem.
This is done most naturally by
using the sum of all ladder diagrams
 $\Gamma_0$ as an effective interaction vertex between $\up$ and $\down$ fermions, or equivalently, a partially dressed pair-propagator.
Diagrammatically, 
\begin{equation}
\includegraphics[width=0.9\columnwidth]{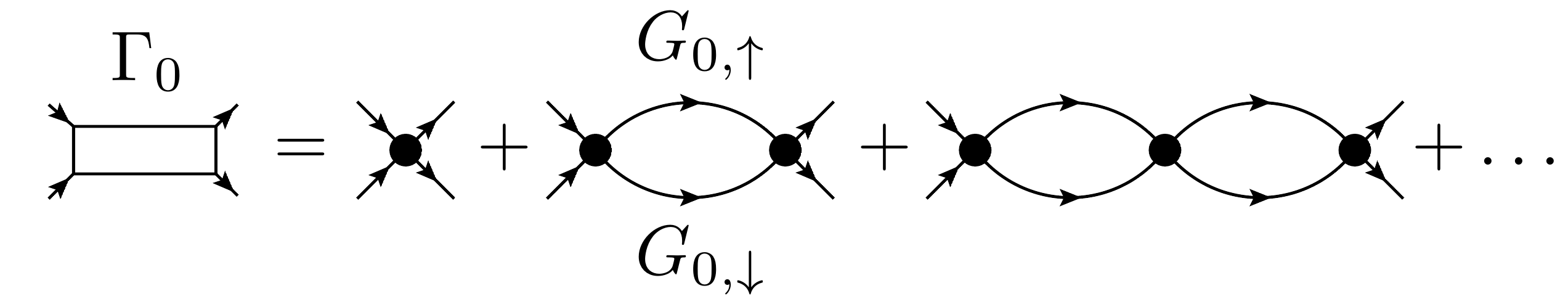} 
\end{equation}
where the $\bullet$ denotes the bare coupling constant and $\Gz$ is the free fermion propagator, given by
$G_{0,\sigma}(p, \omega_{\nu}) = (i \omega_{\nu} + \mu_\sigma - p^2 / 2m)^{-1}$
in the momentum Matsubara-frequency representation.
Here $\sigma\in\{\up,\down\}$, $\mu_\sigma$ is the chemical potential,
$m$ the fermion mass, and
$\omega_{\nu} = (2{\nu}+1)\pi/\beta$ with $\beta=1/(k_B T)$ the inverse temperature.
$\Gaz$ is well-defined for the continuous-space zero-range interaction
(without momentum cutoff) and only depends on the $s$-wave scattering length $a$ (apart from $\mu_\up$, $\mu_\down$, $\beta$, and external momentum-frequency).
The same property holds for higher-order diagrams 
built from $\Gz$ and $\Gaz$.
This ``ladder scheme''
is suited to describe the 
crossover between Fermi and Bose gases,
and its lowest-order approximation
(left diagram in Fig. \ref{fig:diag})
 is widely used~\cite{NozieresSchmittRink,StrinatiUrbanReview,
ChapStrinatiBref}.
 A diagrammatic Monte Carlo algorithm~\cite{VanHoucke1} allows us to stochastically evaluate all 
 Feynman diagrams up to order 9 (see Fig.~\ref{fig:diag}). 

An intensive quantity $Q$, such as pressure or self-energy,
can be formally written as a
diagrammatic series $\sum_{N=0}^\infty a_N$.
Here $a_N$ is a sum of connected diagrams of order $N$
(see Fig.~\ref{fig:diag}).
As we shall see
this diagrammatic series is divergent,
and it is not obvious how to give a meaning to the formal expansion
$Q\overset{?}{=}\sum_{N=0}^\infty a_N$.
To do so, we introduce   a function $Q(z)$ 
whose
Taylor series  is
$\sum_{N=0}^\infty a_N\,z^N$,
and
such that $Q(z=1)$ is the desired exact physical result.
Here $z$ is a formal parameter playing the role of an effective coupling constant.
A non-perturbative 
construction of $Q(z)$ is realised by introducing the action~\cite{ShiftedAction}
\begin{multline}
S^{(z)}
= - 
\int d^3r\,\int_0^\beta d\tau\,
\Big[\,\sum_{\sigma=\up,\down}\bar{\varphi}_\sigma\, G_{0,\sigma}^{-1}\,\varphi_\sigma
+ \bar{\eta}\,\Gaz^{-1}\, \eta
\\- z\, \bar{\eta}\,  \Pi_0 \, \eta + \sqrt{z} ( \bar{\eta}\,\varphi_\down\varphi_\up
+ \bar{\varphi}_\up \bar{\varphi}_\down \eta ) \Big]
\label{eq:shifted_S}
\end{multline}
where $\varphi_\sigma$ are fermionic Grassmann fields, $\eta$ is a bosonic complex field,
and $\Pi_0$ is the particle-particle bubble $(G_{0,\up} G_{0,\down})$,
{\bl 
which cancels out all diagrams containing particle-particle bubbles, as required to avoid double-counting.}
For example, for the pressure we simply have
\be
Q(z) = \underset{\Vr\to\infty}{\lim}\ \frac{1}{\beta\Vr}\,
\ln
\int \Dr \varphi\, \Dr \eta\ e^{-S^{(z)}[\varphi,\eta]}
\ee
with $\Vr$ the volume.

{\it Large-order behavior.}
We now turn to the crucial problem of computing the large-$N$ behavior of $a_N$.
In the pioneering works~\cite{Lipatov,BrezinLipatov}, 
the large-order behavior
for $\phi^4$ theory
 was obtained from a saddle point of the functional integral.
To study the large-order behavior of fermionic theories,
it was found essential to integrate out fermionic fields, which leads to a purely bosonic functional integral,
whose integrand
{\bl $e^{-S_{\rm eff}^{(z)}[\eta]}$}
 can be estimated in the large-field limit using a Thomas-Fermi ({\it i.e.} quasi-local) approximation~\cite{ParisiYukawa,ItzyksonParisiZuber,ItzyksonParisiZuber2,BogomolnyQed,Fry}.
In our problem, this procedure can be justified by showing that
this integrand is an entire function of $z$~\cite{RossiThese,ResonLong2}.
We 
{\bl
find that the bosonic action $S_{\rm eff}^{(z)}[\eta]$ scales as $z^{5/4}\int d^3r d\tau |\eta(\rr,\tau)|^{5/2}$ for large $|\eta|$. 
The saddle-point method then gives}
\begin{multline}
 a_N\underset{N\to\infty}{=}\Gamma(N/5) 
\,A^{-N}
 \\\text{Re}\,\,\text{exp}\left[
i \, 4 \pi N / 5
-U_1e^{i \pi/5}N^{4/5}+O(N^{3/5})\right]
\label{eq:large_n}
\end{multline}
where $U_1 = 5^{1/5} A$ and
 \be
A\coloneqq \frac{1}{\pi^2}\left(\frac{4}{5\,\Gamma(3/4)^8}\right)^{\frac{1}{5}} \min_{\eta}\frac{-\langle\eta|\Gamma_0^{-1}|\eta\rangle}{\left(\int d^3 r\,d\tau\;|\eta(\mathbf{r},\tau)|^{5/2}\right)^{4/5}}.
\label{eq:A}
 \ee

The fact that $a_N$ is of order $(N!)^{1/5}$
immediately implies that the radius of convergence is zero.
This raises a fundamental question: Can the exact physical result still be constructed {\it in a unique way} from the set $\{a_N\}$?

{\it Resummation.}
Given the above asymptotic behavior, it is natural to introduce the generalized Borel transform defined by
 \be
B(z)\coloneqq \sum_{N=0}^\infty \frac{a_N}{\mu_N}\,z^N,\qquad |z|<A
\ee
\be
\mu_N\coloneqq \int_0^\infty 
dt\,
t^4
e^{-t^5-bt^4-ct^3}\,t^N.
\ee
Note that 
$\mu_N \sim \Gamma(N/5) \exp[-b (N/5)^{4/5}]$
for $N\to\infty$.
The corresponding inverse Borel transformation reads
\be
Q_B(z)\coloneqq\int_0^\infty 
dt\,t^4
\,e^{-t^5-bt^4-ct^3}\,B(zt)
\label{eq:invB}
\ee
where $b$ and $c$ are free parameters at this stage.

The answer to the above unicity question is then given by the following theorem {\bl due to Nevanlinna~\cite{Nevanlinna,RamisHouches,Balser}
\footnote{{\bl The parameters $b$ and $c$ in the Borel transform are absent from Nevanlinna's formulation, but we expect that the theorem remains valid in presence of these parameters, since this does not change the leading large-order behavior.}}
\footnote{{\bl
One of the original hypotheses in~\cite{Nevanlinna,Balser}
follows
from
our hypothesis 2 
thanks to Taylor's theorem with Lagrange remainder~\cite{RamisHouches,RossiThese}.}}}.
Let $W\coloneqq \{z\in\mathbb{C}\;|\;0<|z|<R,|\text{arg}\,z|<\pi/10+\epsilon\}$, for some $R>0$ and $\epsilon>0$. If
\begin{enumerate}
  \item{$Q(z)$ is analytical for $z\in W$}
  \item{$\exists \tilde{A}$ and $C$ such that $|d^NQ(z) /dz^N|/N!\le C \,\tilde{A}^{-N}\,(N!)^{1/5}$ for all $N\geq0$ and $z\in W$}
  \item{$a_N=\lim_{z\to 0, z\in W}d^NQ(z) /dz^N/N!$}
  \end{enumerate}
then
\begin{itemize}
  \item
$B(z)$ can be analytically continued for $z\in\mathbb{R}_+$
\item
$\exists R'>0$ such that $Q_B(z)=Q(z)$ for $z\in [0, R']$.
\end{itemize}
The hypotheses of this theorem hold in our situation for the following reasons:
Hypothesis 1 follows from the functional integral representation (\ref{eq:shifted_S}) and the fact that the integrand, after integrating out the fermions,
is an entire function of $z$ that can be bounded in the large-$\eta$ limit using the Thomas-Fermi result.
Hypothesis 2 can be obtained in a similar way to the large-order behavior of $a_N$.
Hypothesis 3 is plausible given that the functional integral for $z\in W$ is absolutely convergent.

\begin{figure}
\includegraphics[width=\columnwidth,trim={0 5cm 0 4.8cm},clip]{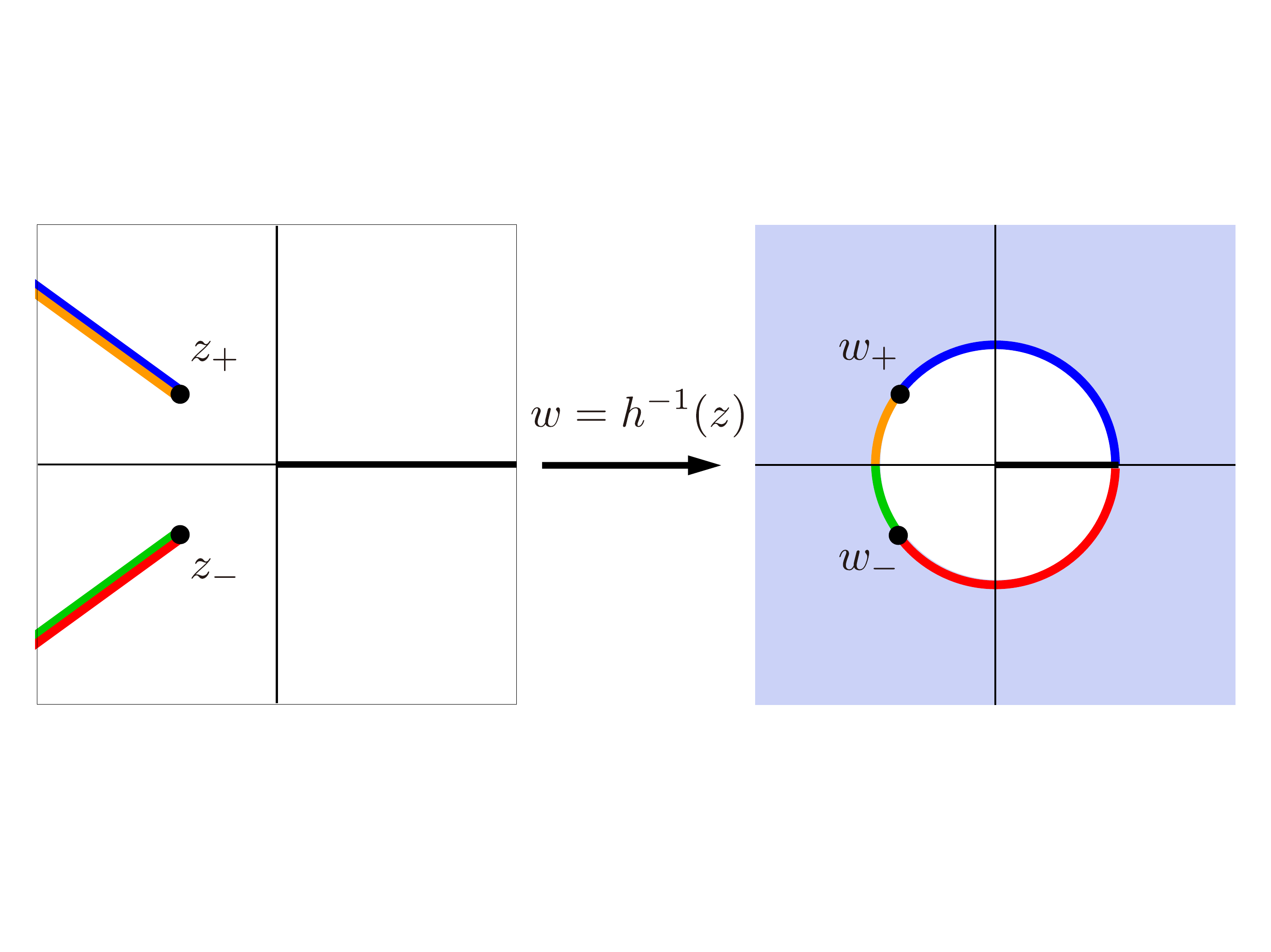}
\caption{Conformal mapping: the singularities of the Borel transform (in color) are mapped onto the unit circle.  The two points $z_{\pm}$ are mapped onto $w_{\pm}$ and 
the real positive axis is mapped onto the segment $[0,1]$.
\label{fig:map}}
\end{figure}

The problem of resummation is thereby reduced to the one of analytical continuation of the Borel transform $B(z)$ to the whole real positive axis.
To this end, it is essential to know the analytical structure of $B(z)$ in the complex $z$ plane.
As follows from the large-order behavior of $a_N$, $B(z)$ has singularities at $z_{\pm} = A\,\exp(\pm i 4\pi/5)$.
We find that $B(z)$ can be analytically continued to $\Dr\coloneqq\mathbb{C}\setminus\{z\in\mathbb{C}\ \big|\ |{\rm arg}\,z|=4\pi/5, \ |z|\geq A \}$
under the condition
$|\theta(b)|< \pi/5$ where
$\theta(b) \coloneqq {\rm arg}[\exp(i\pi/5)\,U_1 - b / 5^{4/5}]$
(note that $b=0$ is not allowed).
This analytical continuation is explicitly realized by 
the one-to-one conformal transformation
$h$ that maps the open unit disc $\Br$ onto $\Dr$ such that $h(0)=0$ and $[h(w)]^* = h(w^*)$, see
Fig.~\ref{fig:map}.
In practice, 
the function
$\tilde{B}(w) \coloneqq B(h(w))$,
{\it a priori} defined for $|h(w)|<A$,
has a Taylor series $\sum_{N=0}^\infty \tilde{B}_N w^N$ that converges for all $w\in\Br$;
$B(z)$ can then be computed as $\tilde{B}(h^{-1}(z)) = \sum_{N=0}^\infty \tilde{B}_N [h^{-1}(z)]^N$ for all $z\in\Dr$.
Substituting this expression into the inverse Borel transform (\ref{eq:invB})
finally yields
$Q(1) = \lim_{N_{\rm max}\to\infty} \sum_{N=0}^{N_{\rm max}} B^{(N_{\rm max})}_N\,a_N$
where the coefficients $B^{(N_{\rm max})}_N$ are computed 
easily thanks to the expression of the conformal mapping
\be
h(w)= \frac{4^{\frac{9}{5}}\,A\,w}{5(1+w)^{2/5}(1-w)^{8/5}}.
\ee

We compute $A$ from \eqref{eq:A} by numerical minimization;
the minimizer $\eta_c$ can be called instanton, or more appropriately soliton since it can be shown to be independent of imaginary time
(it is also rotationally invariant and spatially localized).
We note that $Q(z)$ has the two branch cuts $\{ {\rm arg}\, z = \pm 4\pi/5 \}$,
along which it has discontinuities 
$\sim e^{-(A/|z|)^5}$ for $|z|\to0$
(this follows from the analyticity of $B(z)$ in $\Dr$ and the large-order behavior of $a_N$)
\footnote{For $\phi^4$ theory, 
$Q(z)$ is expected to have only one branch cut,
namely the real negative axis,  with a discontinuity $\sim e^{-A/|z|}$
(the corresponding branch-cut for the Borel transform is $(-\infty, -A]$,
as used in the conformal mapping of~\cite{LeGuillouZinnExposantsPRL}),
and in one-dimensional $\phi^4$ theory there is a clear physical interpretation in terms of the tunneling of a quantum particle 
through the barrier $x^2 + z\,x^4$ for $z<0$,
whose rate is given by the classical action $A/|z|$~\cite{BenderWu1973,ParisiCut,BogomolnyDispersion,ZinnReview,ZinnBook,RivasseauBook,MarinoBook}.
For $\phi^4$ theory in 2 and 3 dimensions, Borel summability was even proven fully rigorously~\cite{MagnenBorel2D,MagnenBorel3D}.}.

\begin{figure}
\includegraphics[width=\columnwidth,clip,trim={1.5cm 2cm 2cm 1cm}]{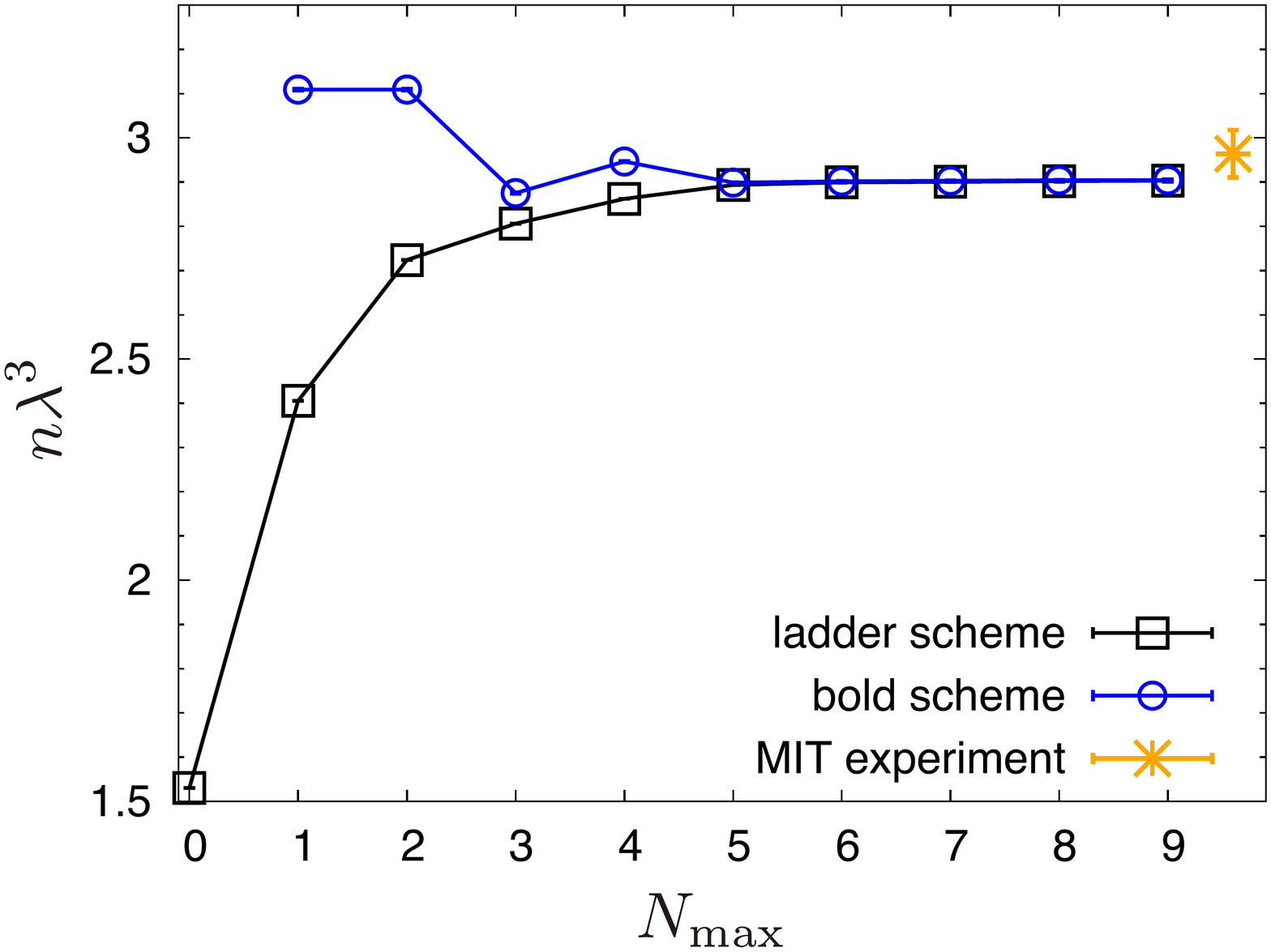}
\caption{Resummed density 
{\it vs.}
maximal diagram order at $\beta\mu = 0$ ($T/T_F \approx 0.6$).
The ladder and bold diagrammatic schemes
agree with each other
and with  experiment.
\label{fig:bm0}}
\end{figure}

\begin{figure}
\includegraphics[width=\columnwidth,clip,trim={1cm 2.2cm 2cm 1cm}]{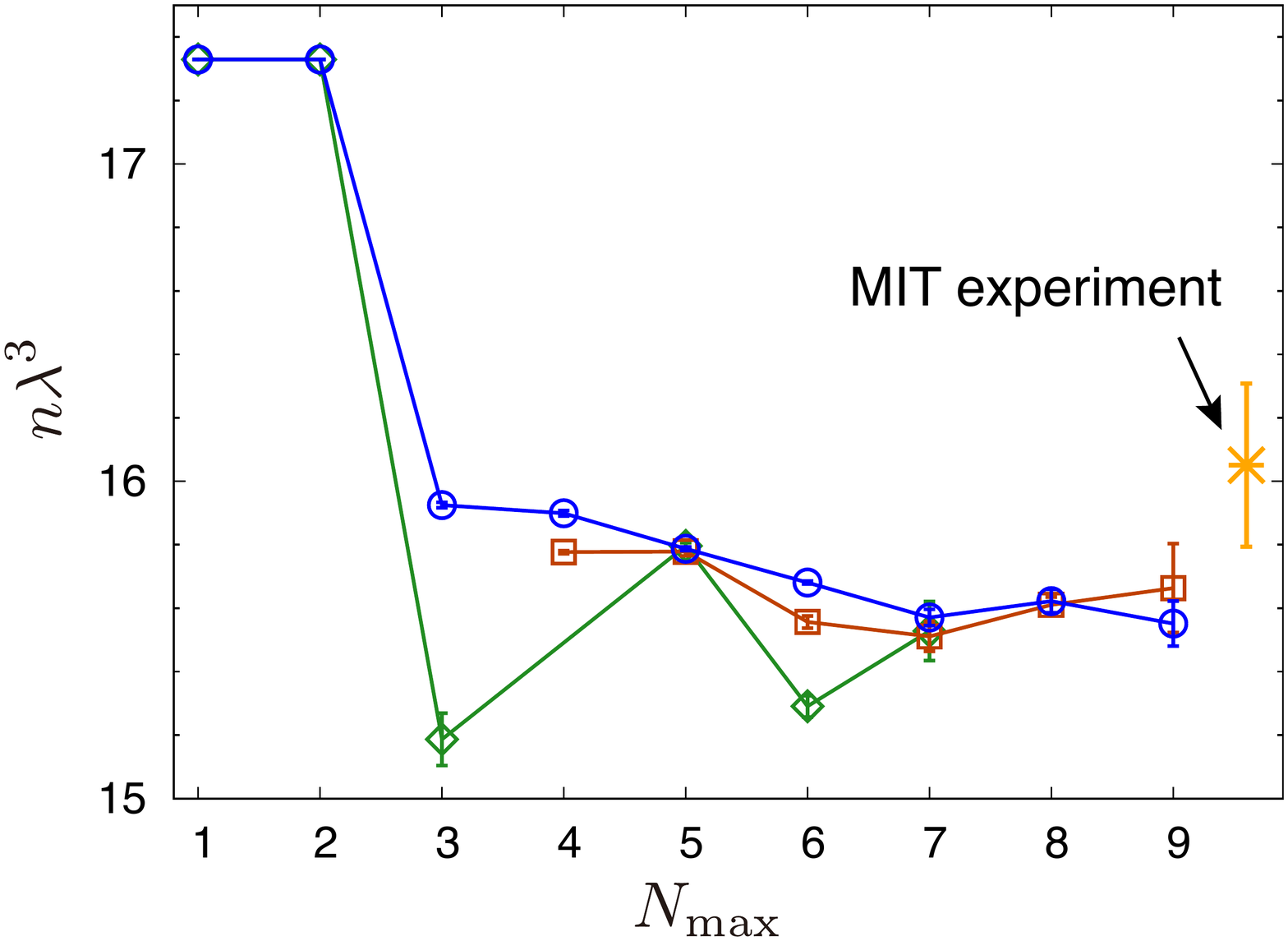}
\caption{Density 
{\it vs.} maximal diagram order at $\beta\mu = 2$ ($T/T_F \approx 0.2$).
The bold diagrammatic series is resummed by three variants of the conformal-Borel transformation (see text).
\label{fig:bm2}}
\end{figure}

{\it Bold scheme.}
In order to access lower temperatures, we turn to the ``bold scheme'' where diagrams are built 
self-consistently from 
fully dressed propagators.
While in the above ladder scheme,
lines and boxes in Fig.~\ref{fig:diag}  denoted
$G_0$ and $\Gamma_0$,
in the bold scheme they denote
 the fully dressed $G$ and $\Gamma$~\cite{BDMC_long}.
Starting from an
action $S^{(z)}_{\rm bold}$ constructed as in~\cite{ShiftedAction},
we find the large-order behavior \eqref{eq:large_n}
modulo the replacement of $\Gamma_0$ by $\Gamma$ in \eqref{eq:A}
and a modified expression for $U_1$~\cite{RossiThese,ResonLong2}.
The justification is less solid than in the ladder scheme, because the integrand
of the purely bosonic functional integral is 
not entire in $z$.
The self-consistent computation is done as follows:
Starting from some initial guess for $G$ and $\Gamma$,
we use the bold diagrammatic Monte Carlo algorithm described in~\cite{BDMC_long}
to compute the (skeleton) diagrammatic series for
the single-particle self-energy $\Sigma$ and the pair self-energy $\Pi$.
We then apply the conformal-Borel resummation procedure to these diagrammatic series.
The resulting resummed $\Sigma$ and $\Pi$ are then plugged into the Dyson equations to obtain new propagators $G$ and $\Gamma$.
This cycle is repeated until convergence~\footnote{Restricting to the lowest-order diagram for $\Sigma$ and $\Pi$ ($N_{\rm max}=1$ in Figs.~\ref{fig:bm0} and \ref{fig:bm2}) is equivalent to the self-consistent T-matrix approximation of Refs.~\cite{Haussmann_PRB,HaussmannZwergerThermo}.}.

We note that 
on approach to the superfluid transition,
$A\to0$ so that the series becomes increasingly hard to resum,
while in the high-temperature limit, $A\to\infty$ (for both ladder and bold schemes) so that the series divergence becomes weaker.

 {\it Numerical results.}
In this paper we focus on 
the central point of the BEC-BCS crossover,
the unitary limit,
where the dimer binding energy vanishes and the scattering length diverges.
This unitary Fermi gas
is strongly correlated since the scattering cross-section is on the order of the squared interparticle distance.
We report results for the Equation of State (EoS) in the normal phase,
 restricting for now to the unpolarized gas, $\mu=\mu_\up=\mu_\down$.
Scale invariance implies than the rescaled density $n\lambda^3$ is a universal function of $\beta\mu$, with $\lambda=\sqrt{2\pi\hbar^2\beta/m}$ the thermal wavelength.

In the moderately degenerate regime, 
we find very good convergence of the series
as a function of the maximal diagram order $N_{\rm max}$ after resummation
by the new conformal-Borel transformation, see Fig.~\ref{fig:bm0}.
The final results for ladder and bold schemes agree within their error bars which are below {\bl $0.1\%$}. 
The value measured at MIT is $2\%$ higher,
a deviation within the experimental uncertainty~\cite{KuEOS}.

Here and in what follows we empirically fixed the free
 parameter $b$ such that $\theta(b) = \pi/10$
(i.e. $b = -5^{4/5} U_1$).
We observed consistent results for different values of
 the free parameter $c$, and we adjusted it to optimise the convergence.
In Fig.~\ref{fig:bm0},
the conformal-Borel transformation was applied to $Q(z)= n(z)$ with $c=12$ for the ladder scheme, and $Q(z)=\Sigma(z)/z$ resp. $\Pi(z)/z$ with $c=10$ for the bold scheme~\footnote{Here, $\Sigma(z)$ and $\Pi(z)$ have the Taylor series
$\sum_{N=1}^\infty \Sigma^{(N)} z^N$
and
$\sum_{N=1}^\infty \Pi^{(N)} z^N$,
where $\Sigma^{(N)}$ and $\Pi^{(N)}$ are defined in~\cite{BDMC_long}.}.
The error bars shown at each $N_{\rm max}$ include the statistical noise coming from the Monte Carlo, and for the bold scheme also the error due to the finite number of iterations.
Our final error bars also include errors due to finite $N_{\rm max}$ and to cutoffs and discretizations in the numerics,
so that all sources of errors are taken into account~\footnote{
Our data for the equation of state is available in the
     Supplemental Material. 
}.

At lower temperatures, the ladder scheme is not applicable
(due to a pole in $\Gamma_0$) 
but we still observe convergence of the bold scheme, as shown in Fig.~\ref{fig:bm2}, where we cross-check three variants of the conformal-Borel resummation:
$Q(z)=\Sigma(z)/z$
resp.
$\Pi(z)/z$
 with $c=13$
(circles),
the same $Q(z)$ with $c=60$
(diamonds),
and 
$Q(z)=\Sigma(z)$
resp.
$\Pi(z)$
with $c=60$
(squares).
Our final result agrees with the MIT measurement up to a $3\%$ deviation
consistent with the experimental uncertainty.

In
the related earlier work~\cite{VanHouckeEOS},
much simpler resummation methods such as the Lindel\"of method were used,
assuming that the diagrammatic series has a non-zero convergence radius.
This assumption is invalidated by the large-order behavior $|a_N| \sim (N!)^{1/5}$ found here.
Hence the results of~\cite{VanHouckeEOS} contained a systematic error. Nevertheless, they deviate from the
 new results reported here by less than 2 \%,
which is likely related to the smallness of
the exponent 1/5.

The sub-factorial scaling $|a_N| \sim (N!)^{1/5}$ also implies that
for a given order $N$,
the sum
 $a_N$ of all diagrams is
 much smaller than the number $\sim N!$ of diagrams.
This is a manifestation of the massive cancellation between different diagrams due to the fermionic sign. 

\begin{figure}
\includegraphics[width=0.9\columnwidth,clip, trim={0.6cm 0.3cm 1.2cm 1.5cm}]{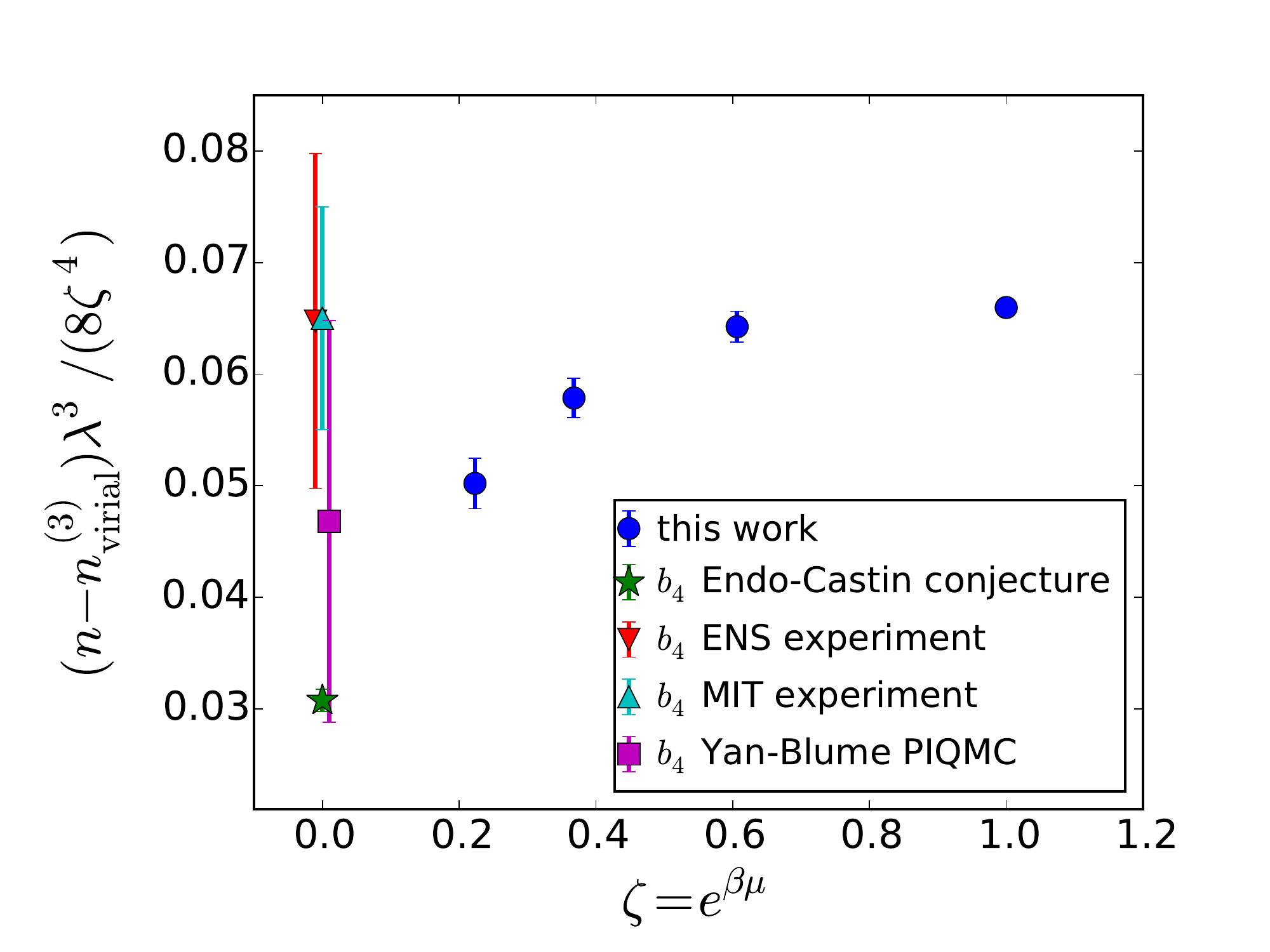}
\caption{Equation of state and 4th virial coefficient:
The difference between
 the density $n$ and its 3rd order virial expansion $n_{\rm virial}^{(3)}$, divided by the appropriate factor, 
must tend to the 4th virial coefficient $b_4$ in the high-temperature limit
 $\zeta\to0$.
\label{fig:b4}}
\end{figure}

Finally we turn to the higher-temperature regime, where our new high-accuracy data sheds light on a controversy.
In the limit $T \gg T_F$, the EoS  admits a virial expansion 
$n_{\rm virial}^{(J)}
\lambda^3 = 2 \sum_{j=0}^J 
j\,b_j\,\zeta^j$
 in powers of the fugacity $\zeta=e^{\beta\mu}$.
The virial coefficient $b_j$ is determined by the $j$-body problem,
and is known exactly for $j=2$~\cite{Ho_virial,RevueTrentoFermions} and $j=3$~\cite{Hu_virial3,GaoEndoCastin}. 
In Figure~\ref{fig:b4} we subtract the known virial-3 result from our EoS data so that the result tends to $b_4$ in the non-degenerate limit $\zeta\to0$.
Accordingly we display at $\zeta=0$ several values reported for $b_4$:
The value obtained by Endo and Castin~\cite{EndoCastinConjecture} 
(based on a physically motivated mathematical conjecture) deviates from the values  reported by experimentalists from
 ENS~\cite{SylEOS} and MIT~\cite{KuEOS}. 
The dedicated Path Integral Quantum Monte Carlo result of Yan and Blume~\cite{YanBlume_b4} 
has an error bar too large to resolve the discrepancy.
Our data suggest that the Endo-Castin result is correct,
but requires sufficiently small $\zeta$ to be extracted,
and correspondingly high accuracy
{\bl to resolve the difference $n-n_{\rm virial}^{(3)} \propto \zeta^4$}
 (at $\zeta\approx0.$2
our error on $n\lambda^3$ is $< 0.01\%$),
while extrapolations from $\zeta \gtrsim 0.6$ lead to the overestimated $b_4$ values reported in~\cite{SylEOS,KuEOS}.
In other words, at $\zeta \approx 0.6$ ($T/T_F\approx 1$) the unitary Fermi gas is still so strongly correlated that it cannot be reduced to a 4-body problem.

In summary,
we found that for the unitary Fermi gas,
a strongly correlated fermion model without small expansion parameter,
diagrammatic series built on partially or fully dressed propagators can be Borel-resummed and yield accurate unbiased results, even though the convergence radius is zero.

How does this relate to other fermionic theories?
For QED, the situation is opposite: 
Large-order behavior and Borel-summability are still open problems~\cite{ItzyksonParisiZuber2,BogomolnyQed,ParisiRenormalon,ZinnBook} 
but no resummation is needed in practice because the coupling constant is small.
QCD combines both difficulties:
It is non-perturbative and probably not Borel-summable~\cite{MarinoBook,RenormalonsQCD},
which calls for new ideas~\cite{DunneUnsalReview}.
The present approach may however be directly generalisable to other continuous-space strongly correlated fermion problems, such as nuclear matter or the electron gas.

\begin{acknowledgements}
{\it Acknowledgements.}
We thank E. Br\'ezin, Y.~Castin, and B.~Svistunov for stimulating discussions, 
and M.~Ku for providing the experimental data of Ref.~\cite{KuEOS}. 
FW was financially supported by
ERC (grants Thermodynamix and Critisup2),
and TO 
by the MEXT HPCI
Strategic Programs for Innovative Research (SPIRE),
the Computational Materials Science Initiative (CMSI)
and Creation of New Functional Devices and High-Performance Materials to Support Next Generation Industries (CDMSI), and by
a Grant-in-Aid for Scientific Research (No.  22104010,
 22340090, 16H06345 and 18K13477) from MEXT, Japan.
Simulations ran on the  clusters `ponyo' at LKB-LPTMC/UPMC, 
`sveta' at UMass,  `blue gene' at IDRIS, `curie' at TGCC, `MesoPSL',
the K computer provided by the RIKEN Advanced Institute for
Computational Science under the HPCI System Research project (project number hp130007, hp140215,
hp150211, hp160201, and hp170263), and at the Supercomputer
Center, Institute for Solid State Physics, University of Tokyo.

\end{acknowledgements}

\bibliography{felix_copy}

\clearpage


{\LARGE Supplemental Material 
}

\vskip 1cm

\beginsupplement

The equation of state can be expressed as density {\it vs.} chemical potential and temperature, $n(\mu,T)$. 
Thanks to scale invariance it reduces to a dimensionless function $n\lambda^3 = f(\beta\mu)$. Our results are given in Table~\ref{tab:dens}. They were obtained using the ladder scheme for $\beta\mu< 0$ and the bold scheme for $\beta\mu \geq 0$.
For cross-check, we also obtained
$n\lambda^3=0.53346(7)$ at $\beta\mu=-1.5$
using
 the bold scheme,
and
$n\lambda^3= 2.903{\bl 3(26)}$
at $\beta\mu = 0$
using the ladder scheme.

We went up to diagram orders
$N_{\rm max}= 10$ at $\beta\mu\leq-1$,
$N_{\rm max}=8$ at 
$\beta\mu=1.5$,
and $N_{\rm max}=9$ in all other cases.

The conformal-Borel transformation was applied to $Q(z) = [n(z)-n(0)]/z$ for the ladder scheme, and to $Q(z)=\Sigma(z)/z$ resp. $\Pi(z)/z$ for the bold scheme. The values used for the free parameter $c$ were
$c=10$ for the bold scheme at $0\leq\beta\mu\leq1$,
$c=15$ at {\bl $\beta\mu=-0.5$ and} $\beta\mu=1.5$,
$c=13$ at $\beta\mu=2$,
$c=20$ at $\beta\mu=-1$,
and $c=12$ in all other cases.

\begin{table}[h!]
\begin{tabular}{|c|c|}
\hline   
$\beta\mu$ & $n\lambda^3$ 
\\
\hline
-1.5 &		0.533477(45)
\\
\hline
-1	&	0.9444{\bl 2}(26)
\\
\hline
-0.5	&     1.673{\bl 5(8)}  
\\
\hline
0 & 		2.9049(26)
\\
\hline
0.5 &    4.821(15)
\\
\hline
1 &  7.54(4)
\\
\hline
1.5 &  11.15(10)
\\
\hline
2 &  15.60(12)
\\
\hline
2.25 &  18.28(22)
\\
\hline
\end{tabular}
\caption{Density equation-of-state.
\label{tab:dens}}
\end{table}

The pressure 
writes
$P(\mu,T)=\int_{-\infty}^\mu d\mu'\,n(\mu',T)$
by the Gibbs-Duhem relation.
In terms of dimensionless functions,
$P(\mu,T)\beta\lambda^3 = F(\beta\mu) 
= \int_{-\infty}^{\beta\mu} dX\,f(X)$.
We evaluate this integral numerically, using an interpolation of the data of Table~\ref{tab:dens}, and the third-order virial expansion for $F(\beta\mu=-1.5)$
(using the fourth-order virial coefficient of~\cite{EndoCastinConjecture} as an error estimate).
The result is given 
in the ancillary file {\sf pressure.txt}.
The error bars include a conservative estimate of the error induced by the interpolation.
{\bl Note that this also yields the energy per unit volume, which equals $3P/2$ by scale invariance~\cite{HoUniv}.}


\end{document}